\documentclass[aps,prd,showpacs,amssymb,superscriptaddress, notitlepage,floats,floatfix,nofootinbib, twocolumn]{revtex4-1}

\usepackage{graphicx}
\usepackage{dcolumn}
\usepackage{bm}
\usepackage{amsmath}
\usepackage{color}
\usepackage{hyperref}

\begin{document}

\title{Cosmological Einstein-Maxwell model  with $g$-essence
}

\author{O. Razina}
\email{olvikraz@gmail.com}
\author{P. Tsyba}
\email{pyotrtsyba@gmail.com}
\author{B. Meirbekov}
\email{beka\_ken@mail.ru}
\author{R. Myrzakulov}
\email{rmyrzakulov@gmail.com}

\affiliation{Eurasian International Center for Theoretical Physics,\\ Department of General $\&$ Theoretical Physics,\\
Eurasian National University, Astana 010008, Kazakhstan}

\date{\today}

\date{\today}

\begin{abstract}
In this paper, we study the model of the late universe with the homogeneous, isotropic and flat Friedmann-Robertson-Walker metric, where the source of the gravitational field is based on the fermion and boson field, with the Maxwell term $F_{\mu\nu}F^{\mu\nu} $ in four dimensions. 
The actuation of the Maxwell term for the Einstein gravity makes it possible to find new approaches to solve the problem of the observed accelerated expansion of the universe. Energy conditions have been obtained and studied. These conditions impose very simple and model-independent restrictions on the behavior of energy density and pressure since they do not require a specific equation of state of matter. To consider the model, the energy conditions NEC, WEC, DEC are  realized, and the SEC condition is violated. The boson and fermion fields are responsible for the accelerated regime at early times, but since the total pressure is tending toward zero for large times, a transition to a decelerated regime occurs. Maxwell field is crucial only in the early times.
\end{abstract}

\maketitle

\section{Introduction}

A few decades ago, cosmologists were studying two quantities: the current rate of expansion of the universe $H_0$ and the deceleration parameter $q_0$. Until the end of the 20th century, it was believed that these two quantities completely determined the final fate of the universe. However, with the discovery of the accelerated expansion of the universe in 1998 , views have completely changed. It turned out that the main role is played by the third quantity -- dark energy. Dark energy (in its various manifestations) finally balanced the energy balance, making the total density of the energy content of the universe equal to the critical density predicted by the theory of inflation. Dark energy has negative pressure and, as a result, causes an accelerated expansion of the universe. Dark energy affects the past and future evolution of the universe. In the case of a cosmological constant, monotonic accelerated expansion occurs. For dynamic forms of dark energy, a number of different scenarios are possible: Big Rip, Big Whimper, Big Decay, Big Crunch, Big Brunch, Big Splat, etc. \cite{Bolotin, Nojiri1}.

Cosmologists still do not fully realize why the expansion of the universe is accelerating.  All doubts can be eliminated by various checks of the kinematics of the expansion, based on physical mechanisms from different areas of physics. In case  the fundamental conclusions about the accelerated expansion of the universe turn out to be erroneous, then the results obtained in different ways will not coincide. At this stage of studying this issue, all the various research methods lead to one conclusion: the universe has entered a period of accelerated expansion \cite{Bolotin, Nojiri1, Nojiri, Bamba2, Bamba3, Bamba, Chiba, Capozziello, Capozziello1, Elizalde2019, Elizalde2018, Elizalde, Dunsby2010, Cognola2008, Cognola2006, Elizalde2004, Sebastiani1, Armendariz1, Armendariz2, Armendariz3, Ribas, Ribas1, Shahalam, Kulnazarov, Myrzakulov, Razina1, Razina2, Putter, Jamil,  Bamba1, Tsyba2015}.

The actuation of the electromagnetic field tensor $F_{\mu\nu}$ (Maxwell term) allows us to find new approaches to solving many problems of cosmology; for instance: the observed the accelerated expansion of the universe.

In this article, we consider the effect of Einstein-Maxwell gravity with g-essence. We have derived the equations of motion. The solution of the model has been plotted. As well, it is determined whether such a model can describe the accelerated expansion of the universe. 

\section{Model Einstein-Maxwell gravity  with g-essence}

We consider the action of Einstein-Maxwell gravity with a g-essence in four dimensions by
\begin{equation} \label{deis}
	S_{fMg}=\frac{1}{16\pi G}\int d^4x \sqrt{-g}\{R-F_{\mu\nu}F^{\mu\nu }+2K(X, Y, \phi, \psi, \bar{\psi}) \}, 
\end{equation}
where the covariant tensor of the electromagnetic field is determined by the derivative of the 4-vector potential in form
\begin{align}\label{fmn1}
	F_{\mu\nu}=\partial_{\mu}A_{\nu}-\partial_{\nu}A_{\mu}
\end{align}
and $K$ is a function of its arguments, $\phi$ is a scalar function, $\psi=(\psi_1, \psi_2, \psi_3, \psi_4)^{T}$ is a fermion function and $\bar{\psi}=\psi^+\gamma^0$ is its conjugate function. Here
\begin {equation}
X=0.5g^{\mu\nu}\nabla_{\mu}\phi\nabla_{\nu}\phi,\quad Y=0.5i[\bar{\psi}\Gamma^{\mu}D_{\mu}\psi-(D_{\mu}\bar{\psi})\Gamma^{\mu}\psi] 
\end{equation}
are the canonical kinetic terms of the scalar and fermion fields, respectively. $\nabla_{\mu}$ and  $ D_{\mu}$ are covariant derivatives. It needs to be noted that the fermion fields are treated here as classical commuting fields.  

The ansatz solution  for the Maxwell term is given by the formula $A_{\mu}$ and is the function from $t$. Then
\begin{align}
	F_{01}=-F_{10}=\dot{A}_1,\ \ F_{02}=-F_{20}=\dot{A}_2,\ \ F_{03}=-F_{30}=\dot{A}_3,  
\end{align}
where the dot on top means the first derivative with respect to time (all other components of $F_{\mu\nu}$ are equivalent to zero).

In general, the ansatz  for the 4-vector potential is
\begin{equation}
A^a_\mu = 
 \begin{cases}
   \phi (t)\delta^a_i, & \mu=i\\
   0, & \mu=0
 \end{cases}. \label{Aam}
\end{equation}

The cosmological principle which asserts that on a huge scale the universe is homogeneous and isotropic. It allows one of a huge number of models describing the universe to choose a narrow class of homogeneous and isotropic models. The most common space-time metric consistent with the cosmological principle is the Friedmann-Robertson-Walker metric
  	\begin{equation}\label{frw}
ds^2=-dt^2+a(t)^2(dx^2+dy^2+dz^2),
\end{equation}
where $a(t)$ is a scale factor of the universe. In some sense, the main task of cosmology is to find the dependence $a(t)$. For this metric, the equation \eqref{fmn1} takes the form
\begin{eqnarray} \label{tensor}
F_{\mu\nu}F^{\mu\nu}=\\2[g^{00}g^{11}(F_{01})^2 +g^{00}g^{22}(F_{02})^2+g^{00}g^{33}(F_{03})^2]=\nonumber\\-\frac{2}{a^2}[(\dot{A}_1)^2+(\dot{A}_2)^2+(\dot{A}_3)^2].\nonumber
\end{eqnarray}
Then the action \eqref{deis} in conjunction with equation \eqref{tensor} can be written in the form 
\begin{equation}\label{fM1}
	S_{fMg}=\frac{1}{8\pi G}\int d^4x \{-3a\dot{a}^2+a[(\dot{A}_1)^2+(\dot{A}_2)^2+(\dot{A}_3)^2]+a^3K \}.
\end{equation}

To find the dependence of the scale factor $a(t)$ on time, equations of motion and an assumption about the material composition of the universe are necessary, which allowing us to construct a pulse energy tensor. The modification of general relativity or the introduction of new components (the $g$ essence, the Maxwell term $F_{\mu\nu}$) changes the dependence of the scale factor on time, but does not affect the relations between the kinematic characteristics. In the case of the FRW metric \eqref{frw}, the equations of motion corresponding to the action \eqref{fM1} could be written as
	\begin{eqnarray}
	3H^2-\rho&=&0,\label{eq1}\\ 
		2\dot{H}+3H^2+p&=&0,\label{feq2}\\
			\ddot{A}_n+H\dot{A}_n&=&0, (n=1,2,3)\label{An}\\
				K_{X}\ddot{\phi}+(\dot{K}_{X}+3HK_{X})\dot{\phi}-K_{\phi}&=&0, \label{kgeq1}\\
	K_{Y}\dot{\psi}+0.5(3HK_{Y}+ \dot{K}_{Y})\psi-i\gamma^0K_{\bar{\psi}}&=&0, \label{deq1}\\ 
K_{Y}\dot{\bar{\psi}}+0.5(3HK_{Y}+\dot{K}_{Y})\bar{\psi}+iK_{\psi}\gamma^{0}&=&0,\label{deq2}\\
	\dot{\rho}+3H(\rho+p)&=&0,\label{eq2}
	\end{eqnarray} 
where the expansion rate of the universe defined by the Hubble parameter $H =\frac{\dot{a}}{a}$ depends on time. The canonical kinetic terms of the scalar and fermion fields have the following form 
\begin{equation}
X=0.5\dot{\phi}^2,\quad  Y=0.5i(\bar{\psi}\gamma^{0}\dot{\psi}-\dot{\bar{\psi}}\gamma^{0}\psi), 
  \end{equation}
 bouth the energy density and pressure of all components represented in the universe at the considered moment of time take the form
\begin{equation}
\rho =2K_{X}X+K_{Y}Y-K+\frac{[(\dot{A}_1)^2+(\dot{A}_2)^2+(\dot{A}_3)^2]}{a^2}, 
\end{equation}
\begin{align}
p=K +\frac{[(\dot{A}_1)^2+(\dot{A}_2)^2+(\dot{A}_3)^2]}{3a^2}.
\end{align}

The equations \eqref{eq1}-\eqref{feq2} are Friedmann equations. The equation \eqref{An} is an equation for the Maxwell term.   The equation \eqref{kgeq1} is the Klein-Gordon equation. The equations \eqref{deq1}-\eqref{deq2} are Dirac equations. The equation \eqref{eq2} is a conservation equation. It follows from the Lorentz invariance of the energy-momentum tensor $T^{\mu}_{\nu,\mu}=0$ which is the first law of thermodynamics for an ideal fluid with constant entropy $dE + pdV = 0$. The equation \eqref{eq2} can be obtained from the Friedmann equations \eqref{eq1}-\eqref{feq2}.

In this article we consider the action of the g-essence of \eqref{fM1} with
\begin {equation}
K=\epsilon X+ \sigma Y- V_{1}(\phi)- V_2(u), 
\end{equation} 
where $ u=\bar{\psi}\psi$, $\epsilon$ and $\sigma$ some constants. Here we can note that $\epsilon=1$ ($\epsilon=-1$) corresponds to the usual (phantom) case, then we can rewrite the system of equations \eqref{eq1}--\eqref{eq2}
\begin{eqnarray}
	3 H^2-\rho &=&0,\label{fi3}\\
	3 H^2+2\dot{H}+p&=&0,\label{fi4}\\
	\ddot{A}_n+H\dot{A}_n&=&0, (n=1,2,3)\label{An1}\\
		\epsilon\ddot{\phi}+3\epsilon H\dot{\phi}+V_{1\phi}&=&0,\label{fi1}\\
	\sigma\dot{\psi}+\frac{3}{2}\sigma H\psi+i V^{'}_2 \gamma^0 \psi&=&0,\label{fi2}\\
	\sigma\dot{\overline{\psi}}+\frac{3}{2}\sigma H\overline{\psi}-i V^{'}_2\overline{\psi} \gamma^0&=&0,\\
		\dot{\rho}+3H(\rho+p)&=&0, \label{fi29}
	\end{eqnarray}
where
\begin{equation}
\rho=0.5\epsilon\dot{\phi}^2+V_1+V_2+\frac{[(\dot{A}_1)^2+(\dot{A}_2)^2+(\dot{A}_3)^2]}{a^2},\label{rho}
\end{equation}
\begin{equation}
p=0.5\epsilon\dot{\phi}^2-V_1-V_2+V^{'}_2u+\frac{[(\dot{A}_1)^2+(\dot{A}_2)^2+(\dot{A}_3)^2]}{3a^2}.  \label{pp}
\end{equation}
Here the prime means the derivative with respect to $u$.

The total energy density of the sources of the gravitational field can be represented as the sum of the three contributions $\rho = \rho_b + \rho_f + \rho_m$, which are associated with the scalar, fermion fields and Maxwell terms, respectively
\begin{equation}
\rho_b=0.5\epsilon\dot{\phi}^2+V_1,\label{rhob} 
\end{equation}
\begin{equation}
\rho_f=V_2,\label{rhof} 
\end{equation}
\begin{equation}
\rho_m=\frac{[(\dot{A}_1)^2+(\dot{A}_2)^2+(\dot{A}_3)^2]}{a^2}.\label{rhom} 
\end{equation}
In the same way, we can represent the total pressure of the sources of the gravitational field as the sum of the pressures $p = p_b + p_f + p_m$ associated with the scalar, fermion fields and Maxwellian terms, respectively
\begin{equation}
p_b=0.5\epsilon\dot{\phi}^2-V_1,  \label{ppb}
\end{equation}
\begin{equation}
p_f=-V_2+V^{'}_2u, \label{ppf}
\end{equation}
\begin{equation}
p_m=\frac{[(\dot{A}_1)^2+(\dot{A}_2)^2+(\dot{A}_3)^2]}{3a^2}.  \label{ppm}
\end{equation}
From \eqref{An1} we obtain the expression for the potential of the 4-vector
\begin{equation}
	\dot{A}_n=\frac{b}{a}, \ \  (n=1, 2, 3) \label{vp}
\end{equation}
where $b=const$.

From the expressions \eqref{Aam} and \eqref{vp} must be
\begin{equation}
	A_n=\phi, \label{vpf}
\end{equation}
\begin{equation}
	a=\frac{b}{\dot{\phi}}. \label{}
\end{equation}
Find all the necessary parameters in terms of $a$
\begin{eqnarray}
  \psi_l&=&\frac{c_l}{a^{1.5}}e^{iD(t)}, \ (l=1,2), \\ 
\psi_k&=&\frac{c_k}{a^{1.5}}e^{-iD(t)}, \ (k=3,4), \\ 
u&=&\dfrac{c}{a^3}, \label{uu}\\
V_1&=& \frac{b^2}{a^2}+V_{10},\label{V1}\\
 V_2&=&-\frac{3b^2}{2a^2}\left(1+\frac{2}{a^2}\right)+6\int{\left(\frac{\ddot{a}}{a^2}-\frac{\dot{a}^2}{a^3}\right)da}, \label{V2}
 \end{eqnarray}
where $c_j$ obeys the following condition $c=|c_{1}|^2+|c_{2}|^2-|c_{3}|^2-|c_{4}|^2$ and
 \begin{equation}
	D=\frac{1}{c}\int{(b^2a+\frac{4b^2}{a}+2a^2\ddot{a}-2a\dot{a}^2)dt}. \label{}
\end{equation}

As an example, consider the case when the scale factor changes according to a power law
\begin{equation}
	a=a_0(t-t_0)^{n}.\label{sf}
\end{equation}
Here $a_0, t_0, n$ are positive constant parameters. For an acceleratingly expanding universe, the condition $n>1$ should be satisfied. The \eqref{fi3}-\eqref{fi29} system has the solution
\begin{eqnarray}
 \phi&=&\frac{b}{a_0(1-n)}(t-t_0)^{1-n}+\phi_0, \\
 \psi_l&=&\frac{c_l}{a^{1.5}}e^{iD(t)}, \ (l=1,2), \\ 
\psi_k&=&\frac{c_k}{a^{1.5}}e^{-iD(t)}, \ (k=3,4), \\ 
u&=&\dfrac{c}{a_0^3(t-t_0)^{3n}}, \label{uu1}
 \end{eqnarray}
where
\begin{eqnarray}
	D= \frac{1}{c}\Bigl[\frac{a_0b^2}{n+1}(t-t_0)^{n+1}+\frac{4b^2}{a_0(1-n)}(t-t_0)^{1-n}\\-\frac{2a_0^3 n}{3n-1}(t-t_0)^{3n-1}\Bigr]+D_0.\nonumber\label{}
\end{eqnarray}
 
For the metric \eqref{frw}  and the  scale factor \eqref{sf}, we obtain the scalar curvature
\begin{align}
	R=6(\dot{H}+2H^2)=\frac{6n(2n-1)}{(t-t_0)^2}.
\end{align}

From \eqref{vp} and the scale factor \eqref{sf} we get an expression for the 4-vector potential depending on time
\begin{equation}
	A_n=\frac{b}{a_0(1-n)}(t-t_0)^{1-n}+\phi_n, \ \  (n=1, 2, 3). \label{vp1} 
\end{equation}

From \eqref{V1} and \eqref{sf} the scalar field potential in terms of $\phi$ is
\begin{equation}
V_1=(1-n)^{\frac{2n}{n-1}}\left(\frac{a_0}{b}\right)^{\frac{2}{n-1}}(\phi-\phi_0)^{\frac{2n}{n-1}}+V_{10}. 
\end{equation}

From \eqref{V2} and \eqref{sf} the potential of the fermion field in terms of $u$ is
\begin{equation}
V_2=-\frac{3b^2}{2}\left(\frac{u}{c}\right)^{\frac{2}{3}}\left(1+2\left(\frac{u}{c}\right)^{\frac{2}{3}}\right)+3n^2a_0^{\frac{2}{n}}\left(\frac{u}{c}\right)^{\frac{2}{3n}}+V_{20}, 
\end{equation}
where $V_{10}$ and $V_ {20}$ are integration constants and for them the relation should be satisfied $V_{10}=-V_{20}$.

Density of energy and pressure are
\begin{eqnarray}
	\rho &=&\frac{3n^2}{(t-t_0)^2},\label{rho1}\\
	p&=&\frac{n(2-3n)}{(t-t_0)^2}.\label{p1}
		\end{eqnarray}

\begin{figure}[htbp]
\vspace*{8pt}
\caption{Energy density \eqref{rho1}, \eqref{rhob1}, \eqref{rhof1}, \eqref{rhom1} as a function of time $t$ ($a_0=2, b=10, n=2, V_{10}=0$). \label{fig:rho1}}
\end{figure}

Energy density  of the scalar, fermion and Maxwell fields

\begin{eqnarray}
\rho_b&=&\frac{3b^2}{2a^{2}_0(t-t_0)^{2n}}+V_{10}, \label{rhob1} \\
\rho_f&=&-\frac{3b^2}{2a^{2}_0(t-t_0)^{2n}}\left(1+\frac{2}{a^{2}_0(t-t_0)^{2n}}\right)+\frac{3n^2}{(t-t_0)^2}-V_{10}, \label{rhof1} \\
\rho_m&=&\frac{3b^2}{a^{4}_0(t-t_0)^{4n}}. \label{rhom1}
\end{eqnarray}

\begin{figure}[pb]
\includegraphics{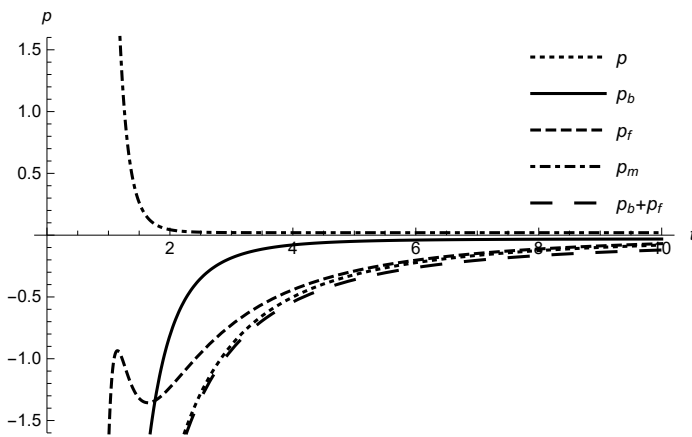}
\vspace*{8pt}
\caption{Pressure \eqref{p1}, \eqref{pb1}, \eqref{pf1}, \eqref{pm1} as a function of time $t$ ($a_0=2, b=10, n=2, V_{10}=0$). \label{fig:p1}}
\end{figure}

Pressure of the scalar, fermion and Maxwell fields

\begin{eqnarray}
p_b&=&-\frac{b^2}{2a^{2}_0(t-t_0)^{2n}}-V_{10}, \label{pb1} \\ 
p_f&=&\frac{b^2}{2a^{2}_0(t-t_0)^{2n}}\left(1-\frac{2}{a^{2}_0(t-t_0)^{2n}}\right)+\frac{n(2-3n)}{(t-t_0)^2}+V_{10}, \label{pf1} \\ 
p_m&=&\frac{b^2}{a^{4}_0(t-t_0)^{4n}}. \label{pm1}
\end{eqnarray}

The transition accelerated can be better understood when we plot the energy and pressure densities as functions of time.  These plots are
shown in Figures \ref{fig:rho1} and  \ref{fig:p1}. 
The pressure of the scalar and  fermionic fields is always negative, while  Maxwell pressure field is positive. The total pressure is always negative, in order to understand the accelerated transition we have to analyze the acceleration field using the equations \eqref{fi4} and \eqref{fi29}, we have $\ddot{a}=-\frac{1}{6}(\rho+3(p_b+p_f+p_m))$ (we have chosen $a = 1$ with $t = t_0$).
At early times Maxwell field was commensurate with the scalar and fermion fields and slowed down the accelerated expansion of the universe. At intermediate times, the density and Maxwell field become small, and the situation arises when $\left|p_b + p_f \right|> \frac{1}{3}\rho + p_m$ which leads to positive acceleration. The scalar and fermion pressures are negative and responsible for the positive acceleration, but they tends to zero with time behaving as dust matter, which corresponds to a decelerated period. We draw attention to the fact that the role of the Maxwell field is important, since with an increase in its value, the transition to the accelerated mode occurs at a later times.

There are a number of different theoretical models that describe the accelerating expansion of the modern universe. For these models, the phenomenological relations $p = \omega\rho$ between the pressure $p$ and the energy density $\rho$ of each of the fluid components, where $\omega$ is the parameter of the equation of state, or for brevity, the parameter of state \cite{Bamba}. A component with a negative $\omega$ corresponds to a dark energy. Modern experiments, including type Ia supernovae \cite{Riess, Perlmutter, Alam}, WMAP \cite{Spergel1, Spergel, Komatsu, Komatsu1}, indicate that at present the dark energy state parameter is close to $-1$. In particular, from the results of observations it follows that with a probability of $0.95$, the value of $\omega$ lies in the interval $\omega = -0.980 \pm 0.053$. From the theoretical point of view, the region mentioned above covers three essentially different cases $\omega>-1$, $\omega=-1$ and $\omega<-1$.
The case $-\frac{1}{3}>\omega>-1$ is realized in quintessence models which are cosmological models with a scalar field. These types of models are quite acceptable, but the question of the origin of this scalar field arises. To comply with astronomical experimental data, the scalar field must be very light and, therefore, not belonging to the field set of the Standard Model. The case $\omega =-1$ is described by using a cosmological constant. This scenario is admissible from a general point of view, but there is a problem associated with the order of magnitude of the cosmological constant \cite{Aref, Armendariz2, Armendariz3}. Phantom matter corresponds to a region of space with the parameter $\omega<-1$, where the scalar field has a negative kinetic energy \cite{Bamba, Chiba}. Although the matter of this form is currently consistent with observations, the origin of such a scalar field with unusual kinetic energy is not clear. On the other hand, it has been recently obtained that the phantom field is not necessarily scalar, it can also have vector or tensor degrees of freedom  \cite{Elizalde, Moon}.

In the model the equation of state parameter is
\begin{equation}
	\omega=\frac{p}{\rho}	=-1+\frac{2}{3n}.
\end{equation}
The obtained state parameter corresponds to modern observational data.

 The components of the energy-momentum tensor $T_{\mu \nu}$ are subject to constraints that follow from the energy conditions.
From these so-called energy conditions, dynamic model-independent constraints on the kinematics of the universe can be obtained. When we are choosing a medium model, these conditions can be transformed into inequalities that impose restrictions on the possible values of pressure and density of the medium. In cosmology, energy conditions are very  important, which in our case have the form \cite{Bolotin}, \cite{Bamba}
\begin{eqnarray}
	NEC &\Rightarrow& \rho+p\geq 0,\label{eu1}\\
	WEC &\Rightarrow& \rho\geq 0, \rho+p\geq 0, \label{eu2}\\
	SEC &\Rightarrow& \rho+3p\geq 0, \rho+p\geq 0, \label{eu3}\\
	DEC &\Rightarrow& \rho\geq 0, -\rho\leq p\leq \rho. \label{eu4}
		\end{eqnarray}
		Here $NEC$, $WEC$, $SEC$ and $DEC$, respectively, are null, weak, strong and dominant energy conditions. These conditions impose very simple and model-independent limitations on the behavior of the energy density and pressure, since they do not require a definite equation of state of matter. Thus, bu using the energy conditions, it is possible to explain the evolution of the universe using general principles \cite{Bolotin, Bamba}.
		
		Since we are considering the model of a flat universe with the FRW metric, we can convert the conditions \eqref{eu1}--\eqref{eu4} into constraints on the deceleration parameter $q$, which is a dimensionless measure of the cosmic acceleration of space expansion
		\begin{eqnarray}
	NEC &\Rightarrow& q\geq -1,\label{uq1}\\
	SEC &\Rightarrow& q\geq 0, \label{uq2}\\
	DEC &\Rightarrow& q\leq 2. \label{uq3}
		\end{eqnarray}
		The condition WEC is always satisfied for arbitrary real $a(t)$.
		
		For $q>0$, the universe expands slowly. For $ q <0 $ the universe expands at an accelerated rate. The condition NEC \eqref{uq1} has a fairly transparent meaning. Accelerated expansion of the universe is possible only in the presence of components with a large negative pressure $p<-\frac{1}{3}\rho$. The SEC energy condition \eqref{uq2} excludes the existence of such components. Hence in this case $q\geq 0$. But the conditions of NEC \eqref{uq1} and DEC \eqref{uq3} are compatible with the condition $p<-\frac{1}{3}\rho$, so they admit regimes in which $q<0$ \cite{Bolotin}.
		
We are considering the model, where the deceleration parameter is
		\begin{equation}
	q=-\frac{\ddot{a}a}{\dot{a}^2}=-1+\frac{1}{n}.\label{q1}
\end{equation}

For the energy density \eqref{rho1}, the pressure \eqref{p1} and the deceleration parameter \eqref{q1}, the energy conditions NEC, WEC, DEC are fulfilled and the SEC condition is not fulfilled. In equation \eqref{q1} $q<0$ by $n>1$ therefore, our model describes the accelerated expansion of the universe.

The jerk parameter $ j $ characterizes the rate of change of the acceleration of the universe expansion
\begin{equation}
	j=\frac{\dddot{a}a^2}{\dot{a}^3}=1-\frac{3}{n}+\frac{2}{n^2}.	\label{j}
\end{equation}
 As can be seen from equation \eqref{j}, the jerk parameter of our model corresponds to the latest observational data	\cite{Riess, Perlmutter, Alam, Spergel1, Spergel, Komatsu, Komatsu1}.

\section{The age of the Universe in Einstein-Maxwell model  with $g$-essence}

The deceleration parameter  as a function of $n$ is therefore given by \eqref{q1}. The condition for acceleration is $q(n) < 0$, thus we have $n>1$. As it can be seen, for $n\rightarrow \infty$ we have $q \rightarrow-1$, that is the universe finally approaches the eternal de Sitter phase with infinite
acceleration, however, the accelerated expansion occurs in $1<n<\infty$.

We consider the accelerated expansion of the universe in the context of Einstein-Maxwell model  with $g$-essence. Lets then start from equation \eqref{q1} and $q(z)=\frac{1+z}{H}\frac{dH}{dz}-1$, find the Hubble parameter as a function of the redshift $z$ as
\begin{equation}\label{Hz}
H(z)=nH_{0}(1+z)^{\frac{1}{n}},
\end{equation}
where $a_{0}/a = 1+z$ with $a_{0}$ and $H_{0}$ being the values of the parameter at the present
epoch. Within the flat universe, dark energy picture, to the Friedmann equations give the expansion rate as \cite{Linder} 
\begin{equation}\label{hh}
H^{2}(z)/H^{2}_{0} = \Omega_{m}(1 + z)^{3} + \delta H^{2}/H^{2}_{0} ,
\end{equation}
where now we encapsulate any modification to the Friedmann equation of general relativity in the last term, $\Omega_{m} = \rho/\rho_{0c}$, $\rho_{0c} = 3H^{2}_{0}$. Also, defining the effective EOS, denoted by $w_{_{\rm eff}}(z)$, as
\begin{equation}\label{weff}
w_{_{\rm eff}}(z) = -1+\frac{1}{3}\frac{d \ln\delta H^{2}}{d \ln(1 + z)} ,
\end{equation}
we can calculate $w_{_{\rm eff}}(z)$ using equations \eqref{Hz}, \eqref{hh} and \eqref{weff} with the result
\begin{equation}\label{kk}
w_{_{\rm eff}}=-1+\frac{1}{3}\frac{\frac{n}{2}(1+z)^{\frac{2}{n}}-3\Omega_{m}(1+z)^{3}}{n^{2}(1+z)^{\frac{2}{n}}-\Omega_{m}(1+z)^{3}}.
\end{equation}
For $n=2$ and $\Omega_{m}=0.33$  we have $w_{_{\rm eff}}\lessapprox-1$, which is the characteristic of one type of dark energy,
the so-called phantom and from equation (\ref{kk}), for $n\rightarrow +\infty$,  we have $w_{_{\rm eff}}\rightarrow-1$.

To continue we consider the age of the universe in Einstein-Maxwell model  with $g$-essence. Thus, the age of the matter dominated Universe in FLRW models is given by
\begin{equation}
t_{0}=\frac{2}{3}\frac{1}{\sqrt{\Omega_{\Lambda}}}H^{-1}_{0}\sinh^{-1} \sqrt{\frac{\Omega_{\Lambda}}{\Omega_{m}}},
\end{equation}
considering $\Omega_{m}+\Omega_{\Lambda}=1$ and inverse hyperbolic sine we get
\begin{equation}
t_{0}=\frac{2}{3}\frac{1}{\sqrt{1-\Omega_{m}}}H^{-1}_{0}\ln\left[\frac{1+\sqrt{1-\Omega_{m}}}{\sqrt{\Omega_{m}}}\right]
\end{equation}
where $H_{0}^{-1} = 9.8 \times 10^{9}h^{-1}$ years and the dimensionless parameter $h$, according to present data, is
about $0.7$. Hence, in the flat matter dominated universe with $\Omega_{total} = 1$ the age of the universe would
be only $t_{0}=\frac{2}{3}H^{-1}_{0}=9.3$ Gyr. This value, even taking into account the uncertainty in the measurement of $H_0$, contradicts the independent restrictions on the age of the universe $\sim13.5$  Gyr \cite{Rubakov}. We obtain the age of the universe  by taking matter in the Friedmann equations as follows
\begin{equation}
t_{0}=n\frac{1}{\sqrt{1-\Omega_{m}}}H^{-1}_{0}\ln\left[\frac{1+\sqrt{1-\Omega_{m}}}{\sqrt{\Omega_{m}}}\right].
\end{equation}

\begin{figure}[h]
\includegraphics{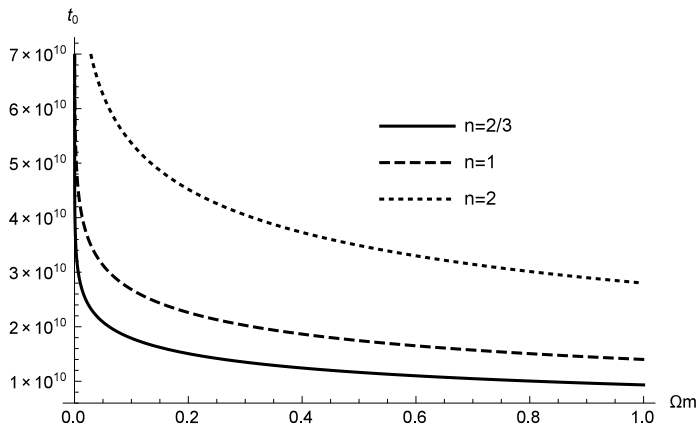}
\vspace*{8pt}
\caption{$t_{0}$ as a function of
$\Omega_{m}$ for $n = 2/3$ (solid line), $n =1$ (dashed line) and $n = 2$ (dotted line). Figure shows that for a fixed value of $\Omega_{m}$ the predicted age of the universe is longer for larger values of $n$.). \label{fig:t0}}
\end{figure}

For a flat,  matter dominated Universe with $\Omega_{m} = 0.33$ and $n =\frac{2}{3}$ we have a prediction for the
age of the Universe of about $13.2$ Gyr. It seems that the age of the universe in our model is longer
than the FLRW model. Figure \ref{fig:t0} shows the behavior of the  age parameter, $t_{0}$, as a function
of $\Omega _{m}$ for different values of $n$.

\section{Conclusion}

Thus, we have considered the model of Einstein-Maxwell gravity with  $g$ essence in four dimensions together with a homogeneous, isotropic and flat Friedman-Robertson-Walker universe. For this model, we have found the power solution, reconstructed the scalar and fermion potentials, and studied the energy conditions. Also, the energy conditions NEC, WEC, DEC are fulfilled and the SEC condition is not fulfilled for this model. The parameter of the equation of state $\omega$, the deceleration parameter $q$ and the jerk parameter $j$ are found, the values correspond to the accelerated expansion of the universe for $n>1$.

In the model under consideration, bouth boson and fermion fields have a negative pressure, and the Maxwell field has a small positive pressure. In the early epoch, the boson and fermion fields are responsible for the accelerated mode, however, when  the total pressure is tending toward zero at a later time, the transition to the slow mode occurs. Maxwell field is important only in the early era.

Based on the proposed model, we analyzed the age of the universe. For the fixed value of $\Omega_m$, the age of the Universe depends on $n$. According to observational data, the age of the universe should be $\sim 13.5$ Gyr., which corresponds to $n=\frac{2}{3}$, and in our model for the realization of the possibility of accelerated expansion of the universe, the condition $n>1$ must be satisfied. Thus, the age of the universe of the investigated model is more than $\sim 13.5$ Gyr.

The study of various models of cosmological acceleration leads to the same conclusion. The more there are these models, the quicker one of a model will be possible to choose, which is the most adequate one when new observational data would be received.

\section*{Acknowledgments}

The work was carried out with the financial support of the Ministry of Education and Science of the Republic of Kazakhstan, Grant No. 0118RK00935.


\end{document}